\documentstyle[prb,aps,twocolumn,epsf]{revtex}

\begin{document}
\twocolumn[\hsize\textwidth\columnwidth\hsize\csname
@twocolumnfalse\endcsname
\draft\title{Spin dynamics of hole doped Y$_2$BaNiO$_5$}
\author{C.D. Batista, A.A. Aligia and J. Eroles}
\address{Centro At\'{o}mico Bariloche and Instituto Balseiro, Comisi\'{o}n Nacional\\
de Energ{\'{\i }}a At\'{o}mica, 8400 S.C. de Bariloche, Argentina.}
\date{Received \today }
\maketitle

\begin{abstract}
Starting from a multiband Hamiltonian containing the relevant Ni and O
orbitals, we derive an effective Hamiltonian $H_{eff}$ for the low energy
physics of doped Y$_{2}$BaNiO$_{5}.$ For hole doping, $H_{eff}$ describes O
fermions interacting with S=1 Ni spins in a chain, and cannot be further
reduced to a simple one-band model. Using numerical techniques, we obtain a
dynamical spin structure factor with weight inside the Haldane gap. The
nature of these low-energy excitations is identified and the emerging
physical picture is consistent with most of the experimental information in Y%
$_{2-x}$Ca$_{x}$BaNiO$_{5}$. \newline

models.\newline
studies, etc).\newline
\end{abstract}
\vskip2pc]

\newpage

The properties of antiferromagnetic (AFM) chains have been a major topic in
the study of low dimensional magnetism. Haldane's conjecture about the
existence of a gap in spin $S$=1 one dimensional (1D) Heisenberg systems 
\cite{hal} has inspired a considerable amount of research \cite{aff,buy,whi}%
. Recently, there has been wide interest in Y$_{2}$BaNiO$_{5}$, a near ideal
realization of the $S$=1 AFM Heisenberg chain \cite{xu}, and the effects of
different dopants on specific heat \cite{ram} and magnetic properties \cite
{dit,koj}. While replacement of Ni$^{2+}$ ($S$=1) by non-magnetic Zn$^{2+}$
or Mg$^{+2}$ simply cuts the spin chains, the replacement of off-chain Y$%
^{3+}$ by Ca$^{2+}$ introduces holes primarily on oxygen sites \cite{dit}.
In the last case, resistivity decreases by several orders of magnitude, and
inelastic neutron scattering reveals new states inside the Haldane gap \cite
{dit}. In addition, the magnetic behavior of the Ca doped system, markedly
different from the Mg-doped one, is spin-glass like with a characteristic
temperature $T_{g}\sim 2K$.\cite{koj}

It has been shown that spin models with localized site or bond impurities
lead to in-gap states for certain parameters \cite{lu,sor}. While the
results are interesting, these models can only be valid for completely
localized added holes, which seems not to be the case in Y$_{2-x}$Ca$_{x}$%
BaNiO$_{5}$ \cite{dit}. Models for mobile holes have been studied \cite
{pen,dag}, but with effective hole hopping $t$ more than one order of
magnitude smaller than realistic values \cite{note0}. In Ref.\cite{dag}, the
essential physics is reduced to that of a weakly mobile $S$=1/2 spin
(representing a ''Zhang-Rice '' doublet \cite{dag,bal}) with exchange
interaction $J^{\prime }=0$ with its nearest-neighbor (NN) Ni spins. In the
static limit, this model is known to lead to a bound state inside the
Haldane gap when $J^{\prime }$ is weak \cite{lu,sor}. However $J^{\prime }$
results strong and ferromagnetic (FM) \cite{note} and, in agreement with
Ref. \cite{sor}, we find no bound states in this static case. Thus, the
origin of the observed states inside the Haldane gap in Y$_{2-x}$Ca$_{x}$%
BaNiO$_{5}$ \cite{dit} is not yet clear.

In this Letter, we start from a multiband Hamiltonian $H_{mb}$ for NiO$_{5}$
linear systems, containing all essential orbitals and interactions. An
effective low-energy Hamiltonian $H_{eff}$ is derived. By exact
diagonalization of $H_{eff}$, we provide an explanation of the observed
inelastic neutron spectrum \cite{dit}. In addition, our results are
consistent with the measured gap, x-ray absorption spectrum \cite{dit},
magnitude of the intra- and inter-chain exchange \cite{xu}, magnetic
susceptibility and muon spin relaxation \cite{koj}.

The starting Hamiltonian is the extension to the 1D NiO$_{5}$ system of the
one used by van Elp {\it et al. }\cite{elp} for a NiO$_{6}$ cluster,
restricting the basis of Ni orbitals to the $3d_{3z^{2}-r^{2}}\;$and $%
3d_{x^{2}-y^{2}}$ ($z$ is the direction along the chain):

\begin{eqnarray}
H_{mb} &=&\sum_{i\alpha }\epsilon _{i}c_{i\alpha \sigma }^{\dagger
}c_{i\alpha \sigma }+\sum_{i\neq j\alpha \sigma }t_{ij}^{\alpha \beta
}c_{i\alpha \sigma }^{\dagger }c_{j\beta \sigma } \nonumber\\
&& +\sum_{i\alpha }(U_{i}+\frac{J_{i}^{H}}{2})n_{i\alpha \uparrow }n_{i\alpha \downarrow}\nonumber\\
&& +\sum_{i\alpha \neq \beta \sigma \sigma ^{\prime }}(U_{i}-\frac{3J_{i}^{H}}{4})n_{i\alpha \sigma }n_{i\beta \sigma ^{\prime }}  \nonumber \\
&& +\sum_{i\alpha \neq \beta }J_{i}^{H}(c_{i\alpha \uparrow }^{\dagger
}c_{i\alpha \downarrow }^{\dagger }c_{i\beta \downarrow }c_{i\beta \uparrow
}/2-{\bf S}_{i\alpha }\cdot {\bf S}_{i\beta })\nonumber\\
&& +\sum_{i\neq j\alpha \beta  
\sigma \sigma ^{\prime }}U_{ij}n_{i\alpha \sigma }n_{j\beta \sigma ^{\prime
}},
\end{eqnarray}
where $c_{i\alpha \sigma }^{\dagger }$ creates a hole with spin $\sigma $ on
the orbital $\alpha $ of site $i$. The interactions included are the on-site
Ni repulsion $U_{d}$, Ni Hund interaction $J^{H}$ (for O sites $J_{i}^{H}=0$%
), O on-site repulsion $U_{p}$ and Ni-O NN repulsion $U_{pd}$, for the six O
atoms surrounding a Ni one. For simplicity, the basis of O orbitals
considered is reduced to the $2p_{z},$ and the two linear combinations of $%
2p $ orbitals lying in the $xy$ plane having optimum hybridization with
their NN Ni orbitals (the amount of other O $2p$ states, in the low energy
manifold, is of the order of 1\%). From atomic data we obtain $J^{H}=1.6$eV.
From Ref. \cite{elp} $U_{d}=U(^{2}E_{g})=10$eV. The values of $U_{p}=4$eV
and $U_{pd}=1.2$eV were taken from Ref. \cite{hyb}. The hopping parameters
were taken from Ref. \cite{elp}, scaling the Ni-O (O-O) hoppings with an $%
r^{-7/2}$ $(r^{-2})$ law, and multiplying them by a common factor $f$.
Finally, $f$ and $\Delta ,$ the energy necessary to take a hole from Ni$%
^{2+} $ and put it in any of the six nearest O$^{2-}$ ions, were chosen to
give $J=0.023$eV and a charge transfer gap $E_{g}=1.84$eV \cite{note3} in
agreement with experiments \cite{dit}.

The resulting values $f=0.85$ (implying a $3d_{3z^{2}-r^{2}}-2p_{z}$ hopping
of 1.61eV ) and $\Delta =7.4$eV are reasonable taking into account that $%
\Delta $ should increase to the left of the periodic table, and $\Delta
=\epsilon _{p}-\epsilon _{d}+U_{pd}\sim 4.8$eV in the cuprates \cite{hyb}.
To gain further confidence on the resulting parameters, we have compared
other two quantities with experiment. Solving exactly $H_{mb}$, including
the repulsion $V_{d}$ $(V_{p})$ between a Ni $2p_{3/2}$ core hole and a Ni $%
3d$ (NN O $2p$) hole in a NiO$_{6}$ cluster, we have calculated the shift
towards lower energy in the Ni $L_{3}$ x-ray absorption spectrum in going
from polarization along the chain to a perpendicular one. Taking $%
V_{d}-V_{p} $=10eV \cite{hyb} we obtain $0.80$eV$,$ while the experimental
shift is $0.91 $eV \cite{dit}. We have also calculated the exchange along $b$
direction (perpendicular to the chains) $J_{b}$, starting from the exact
solution of two NiO$_{6}$ clusters, and treating the hoppings connecting the
two clusters with the cell-perturbation method \cite{sev}. We obtain $%
J_{b}=0.173K=6.4\times 10^{-4}J$, near the upper bound $5\times 10^{-4}J$
estimated in Ref.\cite{xu}.

The low-energy reduction procedure \cite{hyb,sev,ali,bat} which we found
most convenient for the present problem is the one used successfully in Ref.%
\cite{bat} to reproduce low-energy photoemission and magnetic properties of
the three-band model for the cuprates. The {\em form} of the low-energy
effective Hamiltonian $H_{eff}$ is determined by a canonical transformation.
For hole doping $H_{eff}$ has the form of a spin-fermion $H_{sf}$(or
Kondo-Heisenberg) model as in Ref.\cite{pen}. For electron doping, $H_{eff}$
has the form of a one-band model $H_{1b}$ (Eq. (3) with $J^{\prime }=0$ and $%
t=-0.351eV)$. The procedure eliminates linear terms in the Ni-O hopping,
retaining second order contributions and the fourth-order exchange $J$.
However, the {\em values} of the parameters are determined fitting the
energy levels of $H_{sf}$ to the corresponding ones of $H_{mb}$ in
conveniently chosen clusters (Ni$_{2}$O$_{11}$ for electron-doped and NiO$%
_{6}$ for hole-doped systems). $H_{sf}$\ can be written as:

\begin{eqnarray}
H_{sf} &=&\sum_{i\delta \sigma }p_{i+\delta \sigma }^{\dagger }\;p_{i-\delta
\sigma }\left[ \left( t_{1}+t_{2}\right) \left( {\bf S}_{i}\cdot {\bf \Sigma 
}_{i-\delta }+\frac{1}{2}\right) -t_{2}\right] \nonumber\\
&&+J_{K}\sum_{i\delta }\left( 
{\bf S}_{i}\cdot {\bf \Sigma }_{i+\delta }-\frac{1}{2}\right)  \nonumber \\
&&+\frac{J}{2}\sum_{i\delta }{\bf S}_{i}\cdot {\bf S}_{i+2\delta }+\frac{%
\epsilon }{2}\sum_{i\delta \sigma }p_{i+\delta \sigma }^{\dagger
}\;p_{i+\delta \sigma }
\end{eqnarray}
Here $i$ labels a Ni site, and $i+\delta $ denotes its two NN O atoms along
the chain, $p_{i+\delta \sigma }^{\dagger }$ creates an {\em effective} $%
2p_{z}$ hole (it contains information of other Ni and O orbitals) of spin $%
\sigma $ at site $i+\delta $, and ${\bf S}_{i}$ $({\bf \Sigma }_{i+\delta })$
is an effective spin 1 (1/2) at site$\;i\;(i+\delta )$. We obtain $%
t_{1}=0.64 $eV$,\;t_{2}=0.89$eV$\;$and $J_{K}=1.40$eV$;\;\epsilon =8.97$eV
is irrelevant for the spin dynamics, but determines the charge gap \cite
{note3}.

\begin{figure}
\narrowtext
\epsfxsize=3.5truein
\epsfysize=3.1truein
\vbox{\hskip 0.25truein \vskip 0.15truein
\epsfbox[3 20 582 750]{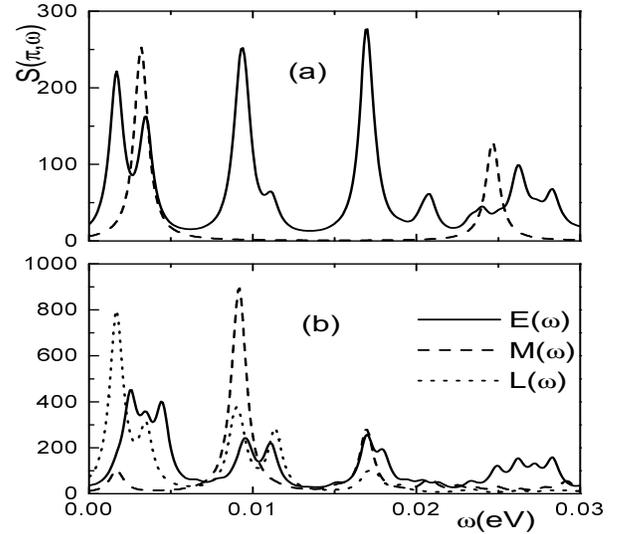}}
\medskip
\caption{(a) Full (dashed) line: dynamic structure factor $S(\omega ,\pi )$%
{\ }$(S(\omega ,\pi )/5{\ })${of O(NiO}$_{5}$)$_{10}$ system {\ with open
boundary conditions and one (zero) added hole. (b) Correlation functions }
$E(\omega ),\;M(\omega )$ and $L(\omega )$ defined in the text.}
\end{figure}

In Fig.1(a) we show the dynamical structure factor $S(q,\omega )=\frac{1}{3}%
\int e^{i\omega t}dt\left\langle {\bf S}_{q}(t).{\bf S}_{-q}(0)\right\rangle 
$ at wave vector $q=\pi $, in a (NiO$_{5}$)$_{10}$O system with one added
hole described by $H_{sf}$, using open boundary conditions (OBC), and
compare it with the result for the undoped system. At first sight, OBC
appears to be an inconvenient choice for studying states inside the Haldane
gap, since an open chain has always low-energy spin excitations associated
with the nearly free spins at the ends \cite{aff,whi,dag}. However, the use
of periodic BC\ leads to spurious low-energy excitations due to the {\em %
frustration}{\it \ }generated by effective FM interactions introduced by the
hole dynamics. This is particularly clear when the spin-spin correlation
between the ends of the open chain is FM. Hence, we used OBC and identified
the excitations related to the end states by the dynamical response of the
system to a staggered magnetic field decaying exponentially along the chain.
For no added holes, $S(\pi ,\omega )$ shows a peak at energy $\omega \sim
J=23$meV corresponding to the Haldane gap (shifted from $\omega \sim 0.4J$
due to the finite size of the chain \cite{whi}), and a peak at $\omega $ $%
\sim $ 3meV corresponding to excitations localized at the ends of the chain.
This is the behavior expected for Zn or Mg doping. $S(\pi ,\omega )$ for one
added hole (representing the Ca doped system) is markedly different. The
Haldane peak shifts to lower energy, its intensity decreases, and
significant spectral weight appears near the middle of the Haldane gap. Both
features are in agreement with experiment \cite{dit}. Among the three peaks
below the Haldane gap, the second one lies nearly at the same energy as that
of the low energy peak of the undoped system. One would then suspect that it
is mainly related with the ends of the chain and it should disappear in the
thermodynamic limit, since Ca doping does not break the chains. This
suspicion is confirmed by inspection of the correlation function $E(\omega
)=\int dt\;e^{i\omega t}\left\langle \frac{1}{2}\left| O(t)O(0)\right| \frac{%
1}{2}\right\rangle $ where $O=\sum_{j}e^{-j(1+i\pi )}S_{j}^{z}$ and $\left|
m\right\rangle $ is the component with $S^{z}=m$ of the $S{\bf =}3/2$ ground
state. $E(\omega )$ measures the response of the system to an excitation
localized at one end of the chain. The result is shown in Fig.1(b). There is
an intense structure around 3meV consisting of three peaks. The first and
the third one are practically absent in $S(\pi ,\omega )$ for symmetry
reasons. The second corresponds precisely to the second peak of $S(\pi
,\omega ).$

What is the origin of the other two peaks below the Haldane gap? Guided by
the wave function of the ground state and the first few excited states, we
have looked for dynamical correlation functions with large intensity at
these peaks to unravel their nature. The peak at 9meV (near the middle of
the Haldane gap) is related to a smaller effective magnetic interaction
between the Ni spins at a distance of one lattice parameter and a half from
the O hole and their NN Ni spins. To check this and since the O hole tends
to be localized near the middle of the chain, we have calculated $M(\omega
)=\int dt\;e^{i\omega t}\left\langle \frac{3}{2}\left|
O_{2}(t)O_{2}(0)\right| \frac{3}{2}\right\rangle $ with $O_{2}=\sum_{\left|
j-c\right| \geq 3/2}(-1)^{j}\;e^{\frac{3}{2}-\left| j-c\right| }S_{j}^{z}$,
where the half-integer $c$ denotes the center of the chain. As shown in
Fig.1(b), $M(\omega )$ has only an intense peak at 9meV, and a smaller
contribution at the Haldane peak. Replacing $O_{2}$ by $%
S_{c+5/2}^{z}-S_{c-5/2}^{z},$ the corresponding dynamical correlation
function $L(\omega )$ has an intense peak only at the energy (around 1.7meV)
of the lowest peak in $S(\pi ,\omega )$, and smaller contributions at the
other low-energy peaks (see Fig.1 (b)). Note instead that the responses to
the end excitation $E(\omega )$ and to the excitation peaked at the middle
of the Haldane gap $M(\omega )$ have negligible contributions at 1.7meV. The
energy of the lowest possible excitation depends non-monotonically on system
size (for example it lies at 0.6meV for O(NiO$_{5})_{8}$), but we find
always an excitation at very low energies. These results suggest the
presence of weak magnetic links near the O hole and the presence of a FM
polaron which is a bounded state between the charge and a local magnetic
excitation. The origin of this polaron, can be described as follows: the
large values of $t_{1},\;t_{2},\;$and $J_{K}$ \ (relative to $J$) tend to
stabilize a doublet between an effective Ni spin at any site $i$ and a hole
in the orbital\ $p_{i+\delta }+p_{i-\delta }$ (''Zhang-Rice'' doublet (ZRD))%
\cite{dag,bal}, and induces a residual FM interaction between the ZRD and
its NN Ni spins \cite{note}. This FM interaction can be interpreted in terms
of a NN attractive interaction between the ZRD\ and a magnetic excitation.
This attractive interaction gives place to a bound state (small polaron)
with a binding energy which is a fraction of $J$. Within this picture, the
effect of the operator $M(\omega )$ is to make a spin excitation {\em inside}
the polaron, while the effect of $L(\omega )$ is to make an excitation in
the spin which is {\em beside} the polaron. It is clear that the AFM
interaction between the latter spin and the one inside the polaron must be
reduced, giving place to a weak link, by the effects of $t_{1},\;t_{2},\;$%
and $J_{K}$ which tend to polarize the spin background. This explains the
origin of the very low-energy excited state, generated by the application of 
$L(\omega )$ to the ground state.

We discuss briefly a further reduction of $H_{sf}$ to a {\em simple} \cite
{note2} one band Hamiltonian $H_{1b}$, since in general, it is successful in
the case of the cuprates \cite{hyb,sev,ali,bat}. The low-energy physics is
dominated by ''Zhang-Rice'' doublets. The overlap of some of these many body
states is too large (1/3 compared with 1/8 in the case of the cuprates \cite
{ali}) rendering cumbersome to obtain $H_{1b}$ as an expansion in powers of
the overlap \cite{ali}. We then construct orthogonal Wannier functions $\pi
_{i}$ at each Ni site, write $H_{sf}$ in this basis using $p_{i+\delta
}=(2/\pi )\sum_{n}(-1)^{n+1}(2n-1)^{-1}\;\pi _{i+n,\sigma }$, calculate $%
P_{1}H_{sf}P_{1}$ ($P_{1}$ is a projector over local orthogonal Zhang-Rice
doublets), and map it into $H_{1b}$ \cite{ali}. Unfortunately, contrary to
the case of the cuprates \cite{ali}, $P_{1}$ does not commute even
approximately with $H_{sf}$ in any limit. Local quadruplets are important in
the low energy physics \cite{note2}. Assuming for simplicity $%
J_{K}=t_{1}+t_{2}$, and including only the two largest terms (besides $J$)
we obtain:

\begin{eqnarray}
H_{1b} &=&\sum_{j\delta }\left[ t\;P_{j,j+2\delta }\;\left( \widetilde{{\bf S%
}}_{j}.{\bf S}_{j+2\delta }+\frac{1}{2}\right) -J^{\prime }\;\widetilde{{\bf %
S}}_{j}.{\bf S}_{j+2\delta }\right] \\
&&+J\sum_{i}{\bf S}_{i}.{\bf S}_{i+1}  \nonumber
\end{eqnarray}
Here $j$ runs over the positions of the Zhang-Rice $S$=1/2 spins $\widetilde{%
{\bf S}}_{j}$ and $P_{j,j\pm 2\delta }\;$permutes $\widetilde{{\bf S}}_{j}\;$%
with its nearest $S$=1 spins ${\bf S}_{j\pm 2\delta }$. The resulting values
of $t$ and $J^{\prime }$ are $t=0.36t_{1}+1.027t_{2}\sim 1eV\sim 40J$,\ and $%
J^{\prime }=0.18J_{K}\sim 0.25eV$, in agreement with simple expectations 
\cite{note0,note}. The $S(\pi ,\omega )$ obtained using $H_{1b}$ is very
different from that of $H_{sf}$ and displays only one peak (around $6$meV)
below the Haldane gap. This result is in fact more similar to the response
of the undoped system.

\begin{figure}
\narrowtext
\epsfxsize=2.5truein
\epsfysize=1.7truein%
\vbox{\hskip 4.75truein \vskip 0.15truein
\epsfbox[0 198 482 780]{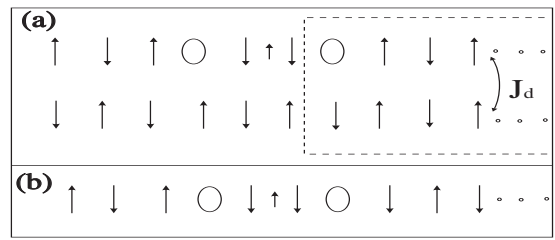}}
\medskip
\caption{(a) Top: schematic representation of the ground state of $H_{ef}$
for one added O hole (indicated by the small up arrow). The dashed line
indicates a region in which the interchain interactions $J_{d}$ are
frustrated due to the change of the interchain short-range magnetic
correlations produced by the O hole. (b)The same for the first excited
state. The Ni states with $S_{{}}^{z}=1,0,-1$ are indicated by $\uparrow
,\;0,\downarrow $ respectively.}
\end{figure}

To shed light onto the underlying physics, it is instructive to solve $%
H_{sf} $ in a Ni$_{3}$O$_{2}$ chain. The ground state has $S$=5/2 and is
nearly degenerate to an $S$=3/2 state. Except for the parity of the wave
functions, the result can be interpreted as if the central static ZRD were
strongly FM coupled to one of the neighboring Ni spins and very weakly FM
coupled to the other one. This quasidegenerancy is not obtained with $H_{1b%
\text{ }}$\ for any set of parameters, indicating that the inclusion of
quadruplets is crucial to have this weak link. This image is confirmed by
the analysis of the dominant terms in the wave function of the ground and
first excited states represented in Fig. 2, and provides a plausible
explanation of the susceptibility measurements of Kojima {\it et al. }\cite
{koj}: at temperatures above the effective exchange of the weak link, the
chains behave as being cut, and a segment between two ZRD's can have zero,
one or both ZRD's FM coupled with it, with probabilities 1/4, 1/2 and 1/4
respectively. If the number of Ni atoms in the segment is even, the spin of
the segment is zero. Then, the corresponding total spin of the segment plus
coupled ZRD's is 0, 1/2, 0 for the three cases respectively. If the number
of sites in the segment is odd, the spin of the segment is equal to 1 and
the corresponding total spin is 1, 3/2 and 2 respectively. This gives an
average $\left\langle S(S+1)\right\rangle =17/8$ in excellent agreement with
the observed Curie constant \cite{koj}.

An important feature of the ground (first excited) state of $H_{sf}$ in the
10-site chain, is that the end spins display FM (AFM) correlations. When the
interchain exchange $J_{b}$ is included, this implies that for 10\% Ca
doping, near five Ni spins have the wrong sign of the magnetic correlations
with neighboring chains, implying a frustration energy of roughly 20$%
J_{b}=3.2K\;$(see Fig. 2). Since this energy is within the order of
magnitude of the randomly (due to disorder \cite{dit} ) distributed lowest
excitation energies, a spin glass behavior with characteristic temperatures
near $3K$ is expected, in agreement with experiment \cite{koj}.

In summary, starting from a multiband Hamiltonian similar to a one used
previously for NiO \cite{elp} but including only $3d$ $e_{g}$ orbitals, we
have derived a low-energy Hamiltonian for doped Y$_{2}$BaNiO$_{5}$, and used
it to understand magnetic properties of the Ca doped system, particularly
the additional states inside the Haldane gap. Contrary to previous
predictions \cite{dag}, we obtain that these states are quite different to
these produced by Zn doping and substantial weight near the middle of the
Haldane gap is obtained only for Ca doping.

Two of us (C.D.B and J.E.) are supported by CONICET, Argentina. A.A.A. is
partially supported by CONICET.




\begin{references}
\bibitem{note0}  On general physical grounds, one expect the interaction
between two $S$=1 Ni spins to be nearly a quarter of that between two $S$%
=1/2 Cu spins in the cuprates, while $t$ should be of the same order of
magnitude. This leads to \cite{hyb,sev,ali}: $J\sim 0.03eV$ (in rough
agreement with experiment \cite{xu,ram,dit}), $t\sim 0.4eV$ in 2D and $t\sim
0.85eV$ in 1D \cite{ali2}.
\bibitem{note}  To understand the origin of $J^{\prime }$, note that when
the Zhang-Rice doublet (ZRD) \cite{dag,bal} has $S_{z}=1/2$ the mean value
of $S_{z}$ in each oxygen is close to $-0.2$ . Then the AFM interaction ($%
J_{K}$) between O and NN Ni spins induces a FM interaction ($J$ $^{\prime }$%
) between the ZRD and its NN Ni spins which is $-0.2J_{K}$ approximately.
\bibitem{note3}  $J$ was determined from the difference between the lowest
triplet and singlet energies of $H_{mb}$ in a Ni$_{2}$O$_{11}$ cluster. The
charge transfer gap is $E_{g}=E(+1)+E(-1)-2E(0)$ where $E(N)$ is the ground
state energy for $N$ added holes using $H_{eff}$ (Eq.(2) for $N\geq 0$
and Eq.(3) with $J^{\prime }=0,$and $t=-0.351eV$ for $N=-1$).
\bibitem{note2}  It is possible that a more complicated one-band
Hamiltonian, including quadruplet corrections \cite{sev}(b,c), three-site
terms, and hoppings and interactions at larger distances, describes
adequately the physics below the Haldane gap. This issue is beyond our scope.
\end{references}
\end{document}